\documentclass[12pt]{article}
\usepackage{graphicx}
\usepackage{natbib} 
\usepackage{url} 
\usepackage{authblk}
\newcommand{\norm}[1]{\left\lVert#1\right\rVert}

\newcommand{\blind}{0}

\usepackage{amssymb}
\usepackage{bm}
\usepackage{amsmath}
\usepackage{booktabs}
\usepackage{multirow}
\usepackage{appendix}
\usepackage{rotating}
\usepackage{mathtools}
\usepackage{graphicx}
\usepackage{algorithm}
\usepackage{algorithmic}

\begin{document}

\def\spacingset#1{\renewcommand{\baselinestretch}%
{#1}\small\normalsize} \spacingset{1}


\if0\blind
{
  \title{\bf Automated Analysis of Experiments using Hierarchical Garrote}
  \author{Wei-Yang Yu}
  \author{V. Roshan Joseph\thanks{Corresponding author: roshan@gatech.edu}}
  \affil{H. Milton Stewart School of Industrial and Systems Engineering, \\
        Georgia Institute of Technology, Atlanta, GA 30332}
  \maketitle
} \fi

\if1\blind
{
  \bigskip
  \bigskip
  \bigskip
  \begin{center}
    {\LARGE\bf Title}
\end{center}
  \medskip
} \fi

\bigskip
\begin{abstract}
In this work, we propose an automatic method for the analysis of experiments that incorporates hierarchical relationships between the experimental variables. We use a modified version of nonnegative garrote method for variable selection which can  incorporate hierarchical relationships. The nonnegative garrote method requires a good initial estimate of the regression parameters for it to work well. To obtain the initial estimate, we use generalized ridge regression with the ridge parameters estimated from a Gaussian process prior placed on the underlying input-output relationship. The proposed method, called HiGarrote, is fast, easy to use, and requires no manual tuning.  Analysis of several real experiments are presented to demonstrate its benefits over the existing methods.



\end{abstract}


\newpage
\spacingset{1.8} 
\section{Introduction}
\label{sec:intro}

Traditionally, the modeling and analysis of experimental data are done using analysis of variance (ANOVA) and regression techniques \citep{box1978statistics, montgomery2017design, wu_experiments_2021}. Regression is more general and the preferred method because some of the factors in the experiments can be continuous and the regression methodology allows us to fit the response surface using smooth polynomial models. When some factors are categorical, they can also be incorporated into the regression methodology by introducing dummy variables. However, because ``orthogonality'' is present in many experimental designs such as full factorial designs, the regression analysis of experimental data is much simpler than that of observational data. The $t$-statistics do not change when an effect is added or removed and therefore, the significant effects can be identified with ease.  However, this simplicity has more or less disappeared as the experimental design methods evolved over time and became more complex. Fractional factorial designs and nonregular designs have become more common in practice where the effects can be aliased, which necessitates the need of more sophisticated regression techniques for the analysis of experiments.

There are certain characteristics of experiments that warrants a different treatment of regression analysis compared to  those commonly used in observational studies. Experiments are usually expensive and therefore, the data are smaller in size. However, experiments are conducted in controlled environments, and therefore, higher order effects such as interactions and nonlinear effects can be entertained in the modeling. The need to model high order effects with small data immediately introduces certain challenges in the regression analysis. First, the total number of effects can exceed and can be much higher than the number of experimental runs. Second, experimenters over time have gained some understanding of the importance of various effects, which are formulated into certain hierarchical principles such as effect hierarchy and effect heredity \citep{wu_experiments_2021}. The effect hierarchy principle states that lower-order effects are more important than higher-order effects.
Under \emph{strong} heredity principle, an interaction is considered active only if both of its parent effects are active; whereas under \emph{weak} heredity, only one of its parent effects needs to be active. These principles are supported by the findings of \citet{li_regularities_2006} and \citet{ockuly_response_2017}, who provide comprehensive meta-analyses of two-level and response surface experiments, respectively. These principles need to be used intelligently to identify the significant effects from the small experimental data.

Analysis of regular designs can be done by first deriving all the aliasing relationships, estimating the aliased effects using regression, and identifying the significant aliased effects using $t$-statistics or half-normal plots and Lenth's method \citep{lenth1989quick}. Then the effect hierarchy and heredity principles are utilized to break the aliases and identify the significant effects. This approach cannot be used for nonregular designs because of the complex aliasing relationships. \cite{hamada_analysis_1992} proposed several forward selection strategies that incorporates the hierarchy and heredity principles. However, these strategies do not explore the complete space of models and therefore, can miss important effects. To search the model space more thoroughly, \citet{chipman_bayesian_1997} proposed a Bayesian variable selection method that integrates the stochastic search variable selection (SSVS) algorithm of \citet{george_variable_1993} with the hierarchical priors described in \citet{chipman_bayesian_1996}.
One major issue with the method is the challenge of specifying priors which contain several tuning parameters. In addition, the approach remains computationally intensive because the SSVS algorithm usually requires a large number of Markov chain Monte Carlo (MCMC) samples to attain convergence.

Several extensions of the regularization-based variable selection methods that preserve effect heredity have been proposed.
\citet{yuan_efficient_2007} presented a fast algorithm by modifying the least angle regression (LARS) technique \citep{efron_least_2004}.
However, their method is not convenient for dealing with situations when one group of factors follows strong heredity and another follows weak heredity.
To provide a more flexible approach, \citet{yuan_structured_2009} incorporated heredity constraints into nonnegative garrote (NG) method of \cite{breiman_better_1995}.
However, NG requires least squares estimates of the parameters to use as initial estimates, which cannot be obtained in fractionated experiments because the number of parameters to  estimate exceeds the number of runs. Extensions of LASSO \citep{tibshirani_regression_1996} can be found in \citet{choi_variable_2010}, \citet{bien_lasso_2013}, and \citet{hao_model_2018}. However, LASSO is not consistent in terms of estimation and variable selection, which is called the oracle property in \citet{fan_variable_2001}.
\citet{vazquez_mixed_2021} built a best-subset selection algorithm that can take any user-specified restriction, including heredity constraints.
Due to the nature of best-subset selection, the algorithm necessitates searching for the optimal model across all-possible model sizes, making it computationally demanding.
More recently, \citet{singh2024factor} proposed an aggregated version of the Gauss-Dantzig Selector \citep{candes_dantzig_2007}.
However, their method is limited to two-level designs only.

The main aim of this article is to develop an automatic method to analyze experimental data which requires little or no manual tuning. To achieve this goal, we will extend the nonnegative garrote method proposed in \cite{yuan_structured_2009} and make it suitable for the analysis of experiments. We chose nonnegative garrote because it possesses the oracle property \citep{zou2006adaptive} and can be automatically tuned \citep{xiong_notes_2010}. The major hurdle in using nonnegative garrote is the non-existence of least squares estimates. \cite{yuan_non-negative_2007} have shown that ridge regression \citep{hoerl_ridge_1970} can be used to obtain initial estimates when the least squares estimates do not exist. However, in this article, we will show that the usual ridge regression estimates do not perform well in the analysis of experiments because the ridge estimates do not obey the hierarchical relationships. A better initial estimate can be obtained using generalized ridge regression, which has a separate ridge parameter for each effect and can be chosen to reflect the hierarchical relationships. However, there are too many ridge parameters, which makes the specification of them challenging.
By exploiting the connection between generalized ridge regression and Bayesian regression, we will use the functionally induced priors proposed in \cite{joseph_bayesian_2006} and \cite{joseph_functionally_2007} to obtain estimates of the ridge parameters which respect the hierarchy and heredity principles. Integrating this with the nonnegative garrote will give us the proposed method, which we call \emph{Hierarchical Garrote} or in short HiGarrote.

The article is organized as follows. In Section \ref{sec:nng}, We will first review the nonnegative garrote method and its limitations. We will then develop our proposed method in Section \ref{sec:methodology}. Several examples are provided in Section \ref{sec:examples} to illustrate the advantages of the proposed method. We conclude the article with some remarks in Section \ref{sec:conc}.

\section{Nonnegative Garrote and Its limitations}
\label{sec:nng}
Suppose that there are $p$ factors $\mathbf{x} = (x_1, x_2, \dots, x_p)'$.
Denote the data as $\{ \mathbf{X}, \mathbf{Y} \} \coloneqq \{ \mathbf{x}_i, y_i \}_{i=1}^{n}$ and let the response $y_i \in \mathbb{R}$, for all $i$. 
Assume that 
\begin{equation}
\label{eq:linear model 1}
    y_i = f(\mathbf{x}_i)+\epsilon_i, \qquad \epsilon_i \sim \mathcal{N}(0,\sigma^2), \qquad \text{for } i = 1, \dots, n,
\end{equation}
where $\epsilon_i$ stands for the random error resulting from uncontrollable factors and measurement error. 
Throughout the paper, we center all variables so that the observed mean is 0 and scale all predictors so that their sample standard deviations are the same.

\subsection{Original Nonnegative Garrote}
\label{subsec:original nng}
Consider approximate the latent function $f$ by a main-effect-only linear model, $$f(\mathbf{x}) \approx \beta_1 x_1 + \beta_2 x_2 + \dots + \beta_p x_p.$$
Let $\bm{\beta} = (\beta_1, \beta_2, \dots, \beta_p)'$. 
The least squares estimator $\hat{\bm{\beta}}^{LS} = (\mathbf{X}'\mathbf{X})^{-1} \mathbf{X}'\mathbf{Y}$.
The original nonnegative garrote (NG) \citep{breiman_better_1995} works by scaling the least squares estimate with  shrinkage factors $\bm{\theta}(M)$ controlled by a tuning parameter $M$. The NG estimate is defined as $\hat{\bm{\beta}}^{NG}(M) = (\hat{\beta}_1^{NG}(M), \dots, \hat{\beta}_{p}^{NG}(M))'$, where $\hat{\beta}_j^{NG}(M) = \hat{\theta}_j(M) \hat{\beta}_j^{LS}$.
For an $M \geq 0$, the shrinkage factors are obtained by solving a quadratic programming problem:
\begin{equation}
\begin{aligned}
\label{eq:original NG}
    \hat{\bm{\theta}}(M)= &\underset{\bm{\theta}}{\text{ argmin}} \; \frac{1}{2} \sum_{i=1}^n \left\{y_i - (\theta_1 \hat{\beta}_1^{LS} x_{i1} + \theta_2 \hat{\beta}_2^{LS} x_{i2} + \dots + \theta_p \hat{\beta}_p^{LS} x_{ip})\right\}^2, \\
    &\text{ subject to } \sum_j \theta_j \leq M \text{ and } \theta_j \geq 0 \; \forall j.
\end{aligned}
\end{equation}
By selecting $M$ suitably, some elements of $\hat{\bm{\beta}}^{NG}(M)$ can be exactly zero.
Since $\hat{\beta}_j^{LS} \neq 0$ with a probability of 1, $x_j$ is considered active if and only if $\hat{\theta}_j(M) > 0$.
In other words, $\hat{\theta}_j(M)$ serves as an indicator of including $x_j$ in the selected model.


\subsection{Nonnegative Garrote with Effect Heredity}
\label{subsec:nng with effect heredity}
In several problems, main effects models are inadequate to approximate the underlying function $f$ in (\ref{eq:linear model 1}). 
To improve the approximation, we can include the quadratic effects and two-factor interactions:
$$f(\mathbf{x}) \approx \beta_1 x_1 + \dots + \beta_p x_p + \beta_{11} x_1^2 + \beta_{12} x_1 x_2 + \dots + \beta_{pp} x_p^2.$$ 
The shrinkage factors in the nonnegative garrote are obtained by 
\begin{equation}
\begin{aligned}
\label{eq:original NG with interactions}
    \hat{\bm{\theta}}(M)= &\underset{\bm{\theta}}{\text{ argmin}} \; \frac{1}{2} \sum_{i=1}^n \left\{y_i - (\theta_1 \hat{\beta}_1^{LS} x_{i1} + \dots + \theta_p \hat{\beta}_p^{LS} x_{ip} + \theta_{11} \hat{\beta}_{11}^{LS} x_{i1}^2 + \dots + \theta_{pp} \hat{\beta}_{pp}^{LS} x_{ip}^2)\right\}^2, \\
    &\text{ subject to } \sum_j \theta_j \leq M \text{ and } \theta_j \geq 0 \;, j = 1, \dots, p, 11, 12, \dots, pp.
\end{aligned}
\end{equation}
In this optimization, it is possible to select a model that includes an interaction term while excluding both of its main effects, that is, $\hat{\theta}_{ij}(M) > 0$ and $\hat{\theta}_{i}(M) = \hat{\theta}_{j}(M) = 0$.
To preserve the effect heredity principle in nonnegative garrote, \citet{yuan_structured_2009} proposed to add constraints that enforce strong or weak heredity among the effects.

Under the strong heredity principle, we must have $\theta_i > 0$ and $\theta_j > 0$ if $\theta_{ij} > 0$.
This can be enforced by requiring $\theta_{ij} \leq \text{min}(\theta_i, \theta_j),$ which is equivalent to two convex constraints 
\begin{equation}
\label{eq:strong heredity constraint}
    \theta_{ij} \leq \theta_i \text{ and } \theta_{ij} \leq \theta_j \;\textrm{for all}\;i,j.
\end{equation}
Thus, if $\hat{\theta}_{ij}(M) > 0$, (\ref{eq:strong heredity constraint}) will guarantee that $\hat{\theta}_i(M) > 0$ and $\hat{\theta}_j(M) > 0$, ensuring that $x_i$ and $x_j$ are part of the chosen model.

Under the weak heredity principle, we must have at least one of $\theta_i$ or $\theta_j$ to be larger than 0 if $\theta_{ij} > 0$.
This can be enforced by requiring $\theta_{ij} \leq \text{max}(\theta_i, \theta_j).$ However, this constraint is nonconvex, and the optimization in (\ref{eq:original NG with interactions}) cannot be solved using quadratic programming.
To simplify the optimization, \citet{yuan_structured_2009} proposed a convex relaxation of the constraint:
\begin{equation}
\label{eq:weak heredity constraint}
    \theta_{ij} \leq \theta_i + \theta_j \;\textrm{for all}\;i,j.
\end{equation}
Thus, if $\hat{\theta}_{ij}(M) > 0$, (\ref{eq:weak heredity constraint}) will ensure that at least one of $\hat{\theta}_i(M) > 0$ or $\hat{\theta}_j(M) > 0$ holds, thereby guaranteeing that at least one of $x_i$ or $x_j$ is included in the selected model.

\subsection{Nonnegative Garrote in Experimental Analysis}
\label{subsec:nng in experimental data}
It is more common in the analysis of experiments \citep{wu_experiments_2021} to approximate $f$ using a linear model with both main effects (linear and quadratic) and their cross product terms (two-factor interactions): 
\begin{equation}
\label{eq:nng model in 2.3}
    f(\mathbf{x}) \approx \beta_1 u_1 + \beta_2 u_2 + \dots + \beta_P u_P,
\end{equation}
where $u_1 = x_1, \dots, u_p = x_p, u_{p+1} = x_1^2, \dots, u_{2p}=x_p^2, \ldots, u_P = x_{p-1}^2x_p^2$, and $P=2p+\binom{2p}{2}$.
For an experiment, the data size $n$ is typically much smaller than the number of parameters $P$, making it impossible to obtain the least squares estimate required in the original nonnegative garrote.
Therefore, \citet{yuan_model_2006} suggested using the ridge regression to obtain an initial estimate in NG.

Let $\mathbf{U}_P$ be an $n \times P$ model matrix.
The ridge regression estimate is given by 
\begin{equation}\label{eq:ridge}
    \hat{\bm{\beta}}^R_P = (\mathbf{U}'_P \mathbf{U}_P + \lambda \mathbf{I}_P)^{-1} \mathbf{U}'_P\mathbf{Y},
\end{equation}
where $\lambda$ is a tuning parameter.
Then, the NG estimate is defined as $\hat{\beta}_j^{NG} = \hat{\theta}_j(M) \hat{\beta}_j^{R}, j = 1, \dots, P$, where the shrinkage factors are obtained as
\begin{equation}
\begin{aligned}
\label{eq:NG in experimental data}
    \hat{\bm{\theta}}(M)= &\underset{\bm{\theta}}{\text{ argmin}} \; \frac{1}{2} \sum_{i=1}^n \left\{y_i - (\theta_1 \hat{\beta}_1^R u_{i1} + \theta_2 \hat{\beta}_2^R u_{i2} + \dots + \theta_P \hat{\beta}_P^R u_{iP})\right\}^2, \\
    &\text{ subject to } \sum_j \theta_j \leq M,\; \theta_j \geq 0 \;, j = 1, \dots, P, \text{ and heredity constraints in \eqref{eq:strong heredity constraint} or \eqref{eq:weak heredity constraint}}.
\end{aligned}
\end{equation}

To tune the hyperparameter $M$, we employ the generalized cross-validation (GCV) criterion introduced by \citet{golub_generalized_1979}, which was extended by \citet{xiong_notes_2010} to handle NG with ridge estimate.
The GCV criterion is defined as
\begin{equation}
\label{eq:GCV formula}
    \begin{aligned}
    \text{GCV}(\hat{\bm{\theta}}(M)) &= \frac{\norm{\mathbf{Y} - \mathbf{U}_P\hat{\bm{\beta}}^{NG}}^2_2}{n(1-d(\hat{\bm{\beta}}^{NG})/n)^2},
\end{aligned}
\end{equation}
where $d(\hat{\bm{\beta}}^{NG})$ is the degrees of freedom of $\hat{\bm{\beta}}^{NG}$.
The degrees of freedom is computed as 
\begin{equation}\label{eq:df}
    d(\hat{\bm{\beta}}^{NG}) = \sum_{i=1}^P \hat{\theta}_i(M) w_i,
\end{equation}
where $w_i$ is the $(i,i)$ entry of the $P \times P$ matrix $(\mathbf{U}_P' \mathbf{U}_P + \lambda \mathbf{I}_P)^{-1} \mathbf{U}_P' \mathbf{U}_P$.
After constructing the solution path across a series of $M \in [0.1, n-1]$, the optimal $M$ is selected by minimizing the GCV criterion,
$$M^{\text{GCV}} = \underset{M}{\text{argmin}} \; \text{GCV} (\hat{\bm{\theta}}(M)).$$
The upper bound of $M$ is set to $n-1$ to ensure that the number of selected terms is not too high compared to the degrees of freedom in the data.

However, we will show that ridge regression estimate is inadequate in experimental data analysis.
Consider a toy example with the 12-run Plackett-Burman (PB) design.
The response is generated by the true model $y = 20A + 10AB + 5AC$.
See Table~\ref{tab:toy example} for the design matrix and data.
We use the R package \texttt{glmnet} to perform the ridge regression and apply the weak heredity constraint when solving the quadratic programming problem \eqref{eq:NG in experimental data}.
This approach  identifies only the effect $A$ with $\hat{\beta}_A^{NG} = 14.938$ and misses the  two interactions $AB$ and $AC$ that are present in the true model.
The issue arises from the underestimation of $\hat{\bm{\beta}}^R_P$.
Due to the presence of numerous independent and irrelevant effects in the model matrix, we found that $\lambda$ is often overestimated, masking the effects of important terms.
In the next section, we will propose an improved initial estimate for NG to overcome the foregoing issue.

\begin{table}[ht!]
\centering
\caption{12-Run Plackett-Burman Design Matrix and Data}
\renewcommand{\arraystretch}{0.5} 
\label{tab:toy example}
\begin{tabular}{crrrrrrrrrrrr}
    \toprule
    \multirow{2}{*}{\text{Run}} & \multicolumn{11}{c}{\text{Factor}} & \multirow{2}{*}{\text{$\mathbf{Y}$}} \\
    \cmidrule(r){2-12}
    \text{} & \text{A} & \text{B} & \text{C} & \text{D} & \text{E} & \text{F} & \text{G} & \text{H} & \text{I} & \text{J} & \text{K}\\
    \midrule
    1 & 1 & 1 & -1 & 1 & 1 & 1 & -1 & -1 & -1 & 1 & -1 & 25 \\
    2 & 1 & -1 & 1 & 1 & 1 & 1 & -1 & -1 & 1 & 1 & -1 & 15 \\
    3 & -1 & 1 & 1 & 1 & 1 & -1 & -1 & -1 & 1 & 1 & 1 & -35 \\
    4 & 1 & 1 & 1 & -1 & -1 & -1 & -1 & 1 & 1 & 1 & -1 & 35 \\
    5 & 1 & 1 & -1 & 1 & -1 & 1 & -1 & 1 & -1 & -1 & 1 & 25 \\
    6 & 1 & -1 & -1 & 1 & 1 & -1 & 1 & -1 & 1 & 1 & 1 & 5 \\
    7 & -1 & -1 & -1 & 1 & -1 & 1 & -1 & 1 & 1 & 1 & 1 & -5 \\
    8 & -1 & 1 & -1 & 1 & -1 & 1 & 1 & 1 & 1 & 1 & -1 & -15 \\
    9 & -1 & -1 & 1 & 1 & 1 & 1 & 1 & 1 & 1 & -1 & -1 & -25 \\
    10 & 1 & -1 & 1 & 1 & -1 & -1 & -1 & 1 & 1 & -1 & -1 & 15 \\
    11 & -1 & 1 & 1 & 1 & 1 & -1 & 1 & -1 & -1 & 1 & 1 & -35 \\
    12 & -1 & -1 & -1 & -1 & -1 & -1 & -1 & -1 & -1 & -1 & -1 & -5 \\
    \bottomrule
\end{tabular}
\end{table}

\section{Methodology}
\label{sec:methodology}
The ridge regression estimate in (\ref{eq:ridge}) is the solution to the following optimization problem:
\begin{equation}\label{eq:ridgeopt}
    \min_{\bm{\beta}_P} (\mathbf{Y}-\mathbf{U}_P\bm{\beta}_P)'(\mathbf{Y}-\mathbf{U}_P\bm{\beta}_P)+\lambda \sum_{i=1}^P\beta_i^2.
\end{equation}
Clearly ridge regression gives equal importance to all the effects, which is not good because the effects are likely to follow the hierarchy and heredity principles. One approach to overcome this issue is to use generalized ridge regression \citep{hoerl_ridge_1970}
\begin{equation}\label{eq:grr}
    \min_{\bm{\beta}_P} (\mathbf{Y}-\mathbf{U}_P\bm{\beta}_P)'(\mathbf{Y}-\mathbf{U}_P\bm{\beta}_P) + \sum_{i=1}^P\lambda_i\beta_i^2,
\end{equation}
where $\lambda_i$'s can be chosen to obey the hierarchical relationships. Although generalized ridge regression is known to perform better than ridge regression in high-dimensional problems \citep{ishwaran2014geometry}, it becomes an arduous task to tune $P>>n$ parameters. Moreover, it is also not clear how to enforce the hierarchical relationships on the $P$ ridge parameters. We will propose an idea to do this.

First note that the generalized ridge regression solution can be viewed as the posterior mean of a Bayesian linear regression with prior:
\begin{equation}\label{eq:prior}
    \bm{\beta}_P \sim \mathcal{N}(\mathbf{0}, \tau^2 \mathbf{R}),
\end{equation}
where $\mathbf{R}$ is a diagonal matrix with $\mathbf{R}_{ii}=\sigma^2/(\tau^2\lambda_i)$. Thus, we only need to determine how to specify the $\mathbf{R}$ matrix in (\ref{eq:prior}). \cite{joseph_bayesian_2006} has proposed an idea to do this by first postulating a Gaussian Process (GP) prior \citep{santner_design_2018} on the underlying function $f(\mathbf{x})$ in (\ref{eq:linear model 1}) and then inducing a prior for the linear model parameters. It is shown in \cite{joseph_bayesian_2006} and \cite{joseph_functionally_2007} that such priors satisfy effect hierarchy and heredity principles under a product correlation structure. Moreover, the prior usually contains only $p$ unknown parameters, which is much less than $P$, and there are well-developed methods to estimate them. This is explained in the next section.

\subsection{Functionally Induced Prior}
\label{subsec:functionally induced prior}
Let the factor $x_j$ have $m_j$ levels in the experiment, $j=1,\ldots,p$.
Consider the model in (\ref{eq:nng model in 2.3}). Although it has only the main effects (linear and quadratic) and their cross product terms (two-factor interactions, 2fi), it is easier to construct the prior for all possible effects.
Thus, we approximate $f$ by a linear model including main effects and all the interactions, i.e.,
\begin{equation}\label{eq:fullmodel}
    f(\mathbf{x}) \approx \sum_{j=0}^{q-1} \beta_j u_j,
\end{equation}
where $q = \prod_{i=1}^p m_i$ and $u_0 = 1.$
To avoid confusion, we let $\bm{\beta}_q = (\beta_0, \beta_1, \dots, \beta_{q-1})'$.

The main effects of the factor $x_j$ can be described by $m_j - 1$ dummy variables, and the interactions can be expressed as the products of these dummy variables.
For example, consider an experiment with two factors each at three levels.
Let $u_1$ and $u_2$ be the main effects of the first factor and $u_3$ and $u_4$ be the main effects of the second factor.
The interaction terms are $u_5 = u_1 u_3$, $u_6 = u_1 u_4$, $u_7 = u_2 u_3$, and $u_8 = u_2 u_4$.
Many popular coding systems can be used to define dummy variables, such as treatment coding, Helmert coding, and orthogonal polynomial coding \citep{wu_experiments_2021, harville1998matrix}.

The idea of functionally induced priors \citep{joseph_bayesian_2006} is to postulate a GP prior for $f$ and use that to induce the prior for $\bm{\beta}_q$.
Thus, assume
\begin{equation}
\label{eq:GP}
    f(\mathbf{x}) \sim \text{GP}(0, \nu^2 \psi),
\end{equation}
where $\nu^2 \psi$ is the covariance function of the Gaussian process defined as $\text{cov} \{ f(\mathbf{x}), f(\mathbf{x+h}) \} = \nu^2 \psi(\mathbf{h})$.
Thus, we have $\mathbf{f} = \mathbf{U}_q \bm{\beta}_q,$
where $\mathbf{f}$ is the vector of function values for the full factorial design and $\mathbf{U}_q$ the $q \times q$ model matrix corresponding to (\ref{eq:fullmodel}).
Note that $\mathbf{f}$ has a multivariate normal distribution with $E(\mathbf{f}) = \mathbf{0}$ and $\text{var}(\mathbf{f}) = \nu^2 \bm{\Psi}$, where $\bm{\Psi}$ is the corresponding correlation matrix.
Since $\bm \beta_q=\mathbf{U}_q^{-1}\mathbf{f}$, we obtain
\begin{equation}
\label{eq:general prior}
\bm{\beta}_q \sim N(\bm{0}, \nu^2 \mathbf{U}^{-1}_q \bm{\Psi} (\mathbf{U}_q^{-1})').
\end{equation}

Assume that the correlation function $\psi$ has a product correlation structure, i.e., 
$$\psi(\mathbf{h}) = \prod_{j=1}^p \psi_j(h_j).$$
Under the product correlation structure, \citet{joseph_functionally_2007} have shown that
\begin{equation}\label{eq:kronecker}
    \text{var}(\bm{\beta}_q) = \nu^2 \bigotimes_{j=1}^{p} \mathbf{U}_j^{-1} \mathbf{\Psi}_j (\mathbf{U}_j^{-1})',
\end{equation}
where $\bigotimes$ denotes the Kronecker product, $\mathbf{U}_j$ is the model matrix for factor $x_j$, and $\bm{\Psi}_j$ the corresponding correlation matrix.
For example, suppose that factor $x_j$ is a three-level factor with levels 1, 2, and 3, the model matrix using orthogonal polynomial coding is
$$\mathbf{U}_j = 
\begin{pmatrix}
    1 & -\sqrt{\frac{3}{2}} & \sqrt{\frac{1}{2}} \\
    1 & 0 & -\sqrt{2}\\
    1 & \sqrt{\frac{3}{2}} & \sqrt{\frac{1}{2}}
\end{pmatrix},
$$
and the correlation matrix is
$$\bm{\Psi}_j = 
\begin{pmatrix}
    1 & \psi_j(1) & \psi_j(2) \\
    \psi_j(1) & 1 & \psi_j(1)\\
    \psi_j(2) & \psi_j(1) & 1
\end{pmatrix}.
$$
The benefit of this structure is that we can select the most appropriate coding system and correlation function for each factor according to our modeling requirements.
Due to its popularity, we use the Gaussian correlation function for each factor \citep{santner_design_2018}: $\psi_j(h_j) = \text{exp}(- h_j^2/\theta_j^2)$, $0 < \theta_j < \infty.$
Let $\rho_j = \exp(-1/\theta_j^2)$.
Then, the correlation function can be written as
\begin{equation}
\label{eq:correlation function rho}
    \psi_j(h_j) = \rho_j^{h_j^2}, \quad 0 < \rho_j < 1.
\end{equation}
Although the form of the correlation function is the same for all factors, the definition of $h_j$ should depend on the type of factors.

For quantitative factors, we define $h_j$ as $|\mathbf{x}_{ij} - \mathbf{x}_{kj}|$, for two runs $i$ and $k$. 
Suppose that there are $p$ quantitative three-level factors and their model matrices are encoded using orthogonal polynomial coding.
Then, (\ref{eq:general prior}) simplifies to \citep{joseph_functionally_2007}:
\begin{equation}
\label{eq:prior for quantitative factor}
    \beta_i \sim \mathcal{N} \left(0, \tau^2 \prod_{j=1}^p r_{j_l}^{l_{ij}} r_{j_q}^{q_{ij}} \right), \quad i = 0, 1, \dots, 3^p-1,
\end{equation}
where $$\tau^2 = \frac{\nu^2}{3^{2p}} \prod_{j=1}^p (3 + 4\rho_j + 2\rho_j^4), \; r_{j_l} = \frac{3-3\rho_j^4}{3 + 4\rho_j + 2\rho_j^4}, \; r_{j_q} = \frac{3-4\rho_j+\rho_j^4}{3 + 4\rho_j + 2\rho_j^4},$$
$l_{ij} = 1$ if $\beta_i$ includes the linear effect of factor $j$ and 0 otherwise, and $q_{ij} = 1$ if $\beta_i$ includes the quadratic effect of factor $j$ and 0 otherwise.
Although the prior for $\bm{\beta}_q$ does includes covariance terms, we do not need them to construct the diagonal matrix $\mathbf{R}$ in (\ref{eq:prior}).

Since $r_{j_l}, \; r_{j_q} \in (0, 1)$, the variance of an interaction effect will be smaller than its parent effects.
In other words, the lower-order effects are likely to be more important than the higher-order effects, and thus the priors satisfy the effect hierarchy principle.
Moreover, since the variance of an interaction effect is proportional to the product of the variances of its parent effects, the priors also satisfy the effect heredity principle.

For qualitative factors where the relative distances between levels cannot be quantified, \citet{joseph_functionally_2007} defines $h_j = 0$ if $\mathbf{x}_{ij} = \mathbf{x}_{kj}$ and 1 otherwise, i.e., $h_j = H(\mathbf{x}_{ij}, \mathbf{x}_{kj})$, where $H(\cdot, \cdot)$ is the Hamming distance.
However, this definition will assign equal importance to all the main effects of $x_j$ even if not all of them are active.
Since the main effects are represented by dummy variables, we define the relative distance for each dummy variable of $x_j$ as $h_{j_{d}} = H(\mathbf{x}_{ij_{d}}, \mathbf{x}_{kj_{d}}), d = 1, \dots, m_j-1$, where the subscript $d$ stands for the $d$th dummy variable.
Then, according to \eqref{eq:correlation function rho}, the form of the corresponding correlation function would be $\psi_{j_{d}}(h_{j_{d}}) = \rho_{j_{d}}^{h_{j_{d}}^2},\; 0 < \rho_{j_{d}} < 1.$

Suppose that there are $p$ qualitative three-level factors and their model matrices are encoded according to Helmert coding.
Then, as shown in the Appendix, the marginals of the prior in \eqref{eq:general prior} will become 
\begin{equation}
\label{eq:prior for qualitative factor}
    \beta_i \sim \mathcal{N} \left(0, \tau^2 \prod_{j=1}^p r_{j_1}^{m_{ij_1}}r_{j_2}^{m_{ij_2}}\right), \quad i = 0, 1, \dots, 3^p-1, 
\end{equation}
where $$\tau^2 = \nu^2 \prod_{j=1}^p (3+2\rho_{j_1}+4\rho_{j_1}\rho_{j_2}), \; r_{j_1} = \frac{3(1-\rho_{j_1})}{3+2\rho_{j_1}+4\rho_{j_1}\rho_{j_2}}, \; 
r_{j_2} = \frac{3+\rho_{j_1} - 4\rho_{j_1}\rho_{j_2}}{3+2\rho_{j_1}+4\rho_{j_1}\rho_{j_2}},
$$
$m_{ij_1} = 1$ if $\beta_i$ includes the first main effect of factor $j$ and 0 otherwise, and $m_{ij_2} = 1$ if $\beta_i$ includes the second main effect of factor $j$ and 0 otherwise.
As in the case of quantitative factors, effect hierarchy and heredity are built-in in these priors.

\subsection{Initial Estimate}
\label{subsec:initial estimator}
Our aim is to obtain an initial estimate of $\bm \beta_P$ from
$\mathbf{Y} =\mathbf{U}_P \bm \beta_P + \epsilon$,
where $\epsilon\sim \mathcal{N}(\bm 0,\sigma^2 \mathbf{I})$ and $\bm \beta_P\sim \mathcal{N}(\mathbf{0}, \tau^2 \mathbf{R})$. Here $\mathbf{R}$ is a diagonal matrix, which is constructed as in (\ref{eq:prior for quantitative factor}) and (\ref{eq:prior for qualitative factor}). The posterior mean of $\bm \beta_P$ is given by
\begin{eqnarray}
    \hat{\bm{\beta}}_P &=& \left(\mathbf{U}_P'\mathbf{U}_P+\frac{\sigma^2}{\tau^2}\mathbf{R}^{-1}\right)^{-1}\mathbf{U}_P'\mathbf{Y}  \nonumber\\
    &=& \tau^2 \mathbf{R} \mathbf{U}_P' \left(\tau^2 \mathbf{U}_P \mathbf{R} \mathbf{U}_P' + \sigma^2 \mathbf{I}_n\right)^{-1} \mathbf{Y}.\label{eq:posterior 1}
\end{eqnarray}

To obtain the initial estimate of $\bm \beta_P$ using \eqref{eq:posterior 1}, several parameters need to be specified, which include $\sigma^2$, $\tau^2$, and $\bm{\rho}$ that define $\mathbf{R}$.
Let $\bm{\rho}=(\bm{\rho}_1', \dots, \bm{\rho}_p')'$ where $\bm{\rho}_j$ can be a vector.
A good estimate of $\sigma^2$ can be obtained if the experiment contains replicates.
Suppose that there are $m$ replicates in each run and let $s_i^2$ be the sample variance for the $i$th run, $i = 1, \dots, n$.
Then $\hat{\sigma}^2 = \sum_{i=1}^n s_i^2 / n$ can be a good estimate of $\sigma^2$.
However, if the experiment is unreplicated, there will be no information to estimate $\sigma^2$.
Instead of assuming that there is no noise, we assume that the noise would contribute at least $1\%$ of the variation in the response to prevent overfitting.
Therefore, we reparameterize \eqref{eq:posterior 1} to
\begin{equation}
\label{eq:posterior 2}
    \hat{\bm{\beta}}_P = \frac{\tau^2}{\nu^2} \mathbf{R} \mathbf{U}_P' \left[ \frac{\tau^2}{\nu^2} \mathbf{U}_P \mathbf{R} \mathbf{U}_P' + \frac{\lambda}{1-\lambda} \mathbf{I}_n\right]^{-1} \mathbf{Y},
\end{equation}
where $\lambda = \frac{\sigma^2}{\sigma^2 + \nu^2}\ge 0.01$.
This form can also be used for replicated experiments as long as we replace $\mathbf{Y}$ with the sample means and divide $\mathbf{I}_n$ by $m$.

\citet{joseph_functionally_2007} have shown that if we use orthogonal coding for each factor,
$$\frac{\tau^2}{\nu^2} = \frac{\prod_{j=1}^p \text{sum}(\mathbf{\Psi}_j)}{q^2},$$
where $\text{sum}(\mathbf{\Psi}_j)$ is the sum of all elements in $\mathbf{\Psi}_j$.
Therefore, we only need to focus on the specifications of $\bm{\rho}$ and $\lambda$.
These hyperparameters can be estimated by minimizing the negative log-likelihood  of $\mathbf{Y}$ \citep{santner_design_2018}.
Thus,
\begin{eqnarray}
    \hat{\lambda},\; \hat{\bm{\rho}} &=& 
\underset{\lambda, \bm{\rho}}{\text{argmin}} \log \hat{\nu^2} + \frac{1}{n} \log \det\left(\bm{\Psi}_n + \frac{\lambda}{1-\lambda} \mathbf{I}_n\right),\label{eq:nlik}\\
\hat{\nu^2} &=& \frac{1}{n} \mathbf{Y}' \left(\bm{\Psi}_n + \frac{\lambda}{1-\lambda} \mathbf{I}_n \right)^{-1} \mathbf{Y},\nonumber
\end{eqnarray}
where $\bm{\Psi}_n$ is the correlation matrix of the $n$ runs.

Different from the usual applications of GP modeling, we are not only interested in the accurate prediction of the output but also the selection of important effects. Therefore, it is very important for us to obtain the global optimum of (\ref{eq:nlik}). This is not easy because when $n$ is small,  several models can fit the data well and therefore, the likelihood can be multi-modal with several local optima. While numerous global optimization algorithms are available for this purpose, we need one that requires only a few function evaluations and is robust to the choice of starting points.
Therefore, we adopt a gradient-based multi-start global optimization strategy.

For numerical stability, we put mild constraints on the hyperparameters: $\rho_i \in [10^{-15}, .999]$ for all $i=1,\ldots,k$, and $\lambda \in [.01, .99]$.
First, we generate a space-filling design with $k+1$ starting points in the feasible region.
Here, we use the MaxPro design in \citet{joseph_maximum_2015}.
Second, for each starting point, we search for the local optimum using a local optimization algorithm.
Here we use the Method of Moving Asymptotes in \citet{svanberg_class_2002}.
Among those local optima, the one with the lowest negative log-likelihood value will be used as the estimate of the hyperparameters.
Importantly, we choose a derivative-based approach not only because gradients help us reduce the number of function evaluations but also because it gives us a more robust result than other derivative-free algorithms.

\subsection{HiGarrote}
Once the hyperparameters are estimated from (\ref{eq:nlik}), the shrinkage factors and the regression coefficients can be obtained as in (\ref{eq:NG in experimental data}) using the initial estimates from (\ref{eq:posterior 2}).
The tuning parameter $M$ in (\ref{eq:NG in experimental data}) can be selected as described in Section~\ref{subsec:nng in experimental data}.
A solution path across a range of $M \in [0.1, 0.3(n-1)]$ is first constructed.
The optimal $M$ is then selected by minimizing the GCV criterion  (\ref{eq:GCV formula}), where the degrees of freedom is calculated as in (\ref{eq:df}) with 
$$w_i=\frac{\tau^2}{\nu^2} \left(\mathbf{R} \mathbf{U}_P' \left[ \frac{\tau^2}{\nu^2} \mathbf{U}_P \mathbf{R} \mathbf{U}_P' + \frac{\lambda}{1-\lambda} \mathbf{I}_n\right]^{-1} \mathbf{U}_P\right)_{ii}.$$
The upper bound of $M$ used in the optimization is based on the findings of \citet{box1986analysis} and \citet[p.~424]{wu_experiments_2021}, which indicate that the proportion of active effects (with intercept term included) typically ranges between 0.13 and 0.27 of the design's run size.

For replicated experiments where a reliable estimate of $\sigma^2$ can be obtained, the shrinkage factors can be directly computed by solving \citep{xiong_notes_2010}:
\begin{equation}
\begin{aligned}
\label{eq:revisit NG for replicated data}
    \hat{\bm{\theta}} = &\underset{\bm{\theta}}{\text{ argmin}} \frac{1}{2} \norm{\mathbf{Y} - \mathbf{U}_P\hat{\bm{\beta}}^{NG}}^2_2 + \hat{\sigma}^2 d(\hat{\bm{\beta}}^{NG}), \\
    &\text{ subject to } \text{heredity constraints in \eqref{eq:strong heredity constraint} or \eqref{eq:weak heredity constraint}}.
\end{aligned}
\end{equation}

Algorithm~\ref{alg:HiGarrote} represents the proposed Hierarchical Garrote (HiGarrote) method. Most of the computational overhead of the method is in step 1 of the algorithm, i.e., the optimization in (\ref{eq:nlik}).

\begin{algorithm}
\caption{HiGarrote}
\label{alg:HiGarrote}
\begin{algorithmic}[1]
    \STATE Estimate $\lambda$ and $\bm{\rho}$ by (\ref{eq:nlik})
    \STATE Obtain initial estimate $\hat{\bm{\beta}}_P$ by (\ref{eq:posterior 2})
    \STATE Obtain $\hat{\bm{\theta}}$ by solving the quadratic programming problem as in (\ref{eq:NG in experimental data}) or (\ref{eq:revisit NG for replicated data})
    \STATE Return $\hat{\bm{\beta}}^{NG}$ by scaling $\hat{\bm{\beta}}_P$ with $\hat{\bm{\theta}}$
\end{algorithmic}
\end{algorithm}

\subsection{Simulation}
To demonstrate the effectiveness of HiGarrote, a simulation study is conducted using the toy example described in Section~\ref{subsec:nng in experimental data}, based on a 12-run PB design.
The response is generated from the model
$$y = 20A + 10AB + 5AC + \epsilon, \qquad \epsilon \overset{iid}{\sim} \mathcal{N}(0, 1).$$
The simulation is repeated 100 times, with new data generated each time.
For each simulated dataset, both HiGarrote and the nonnegative garrote (NG) with ridge estimate \citep{yuan_structured_2009} are fitted using weak heredity.

\begin{figure}[ht!]
    \centering
    \includegraphics[scale=0.15]{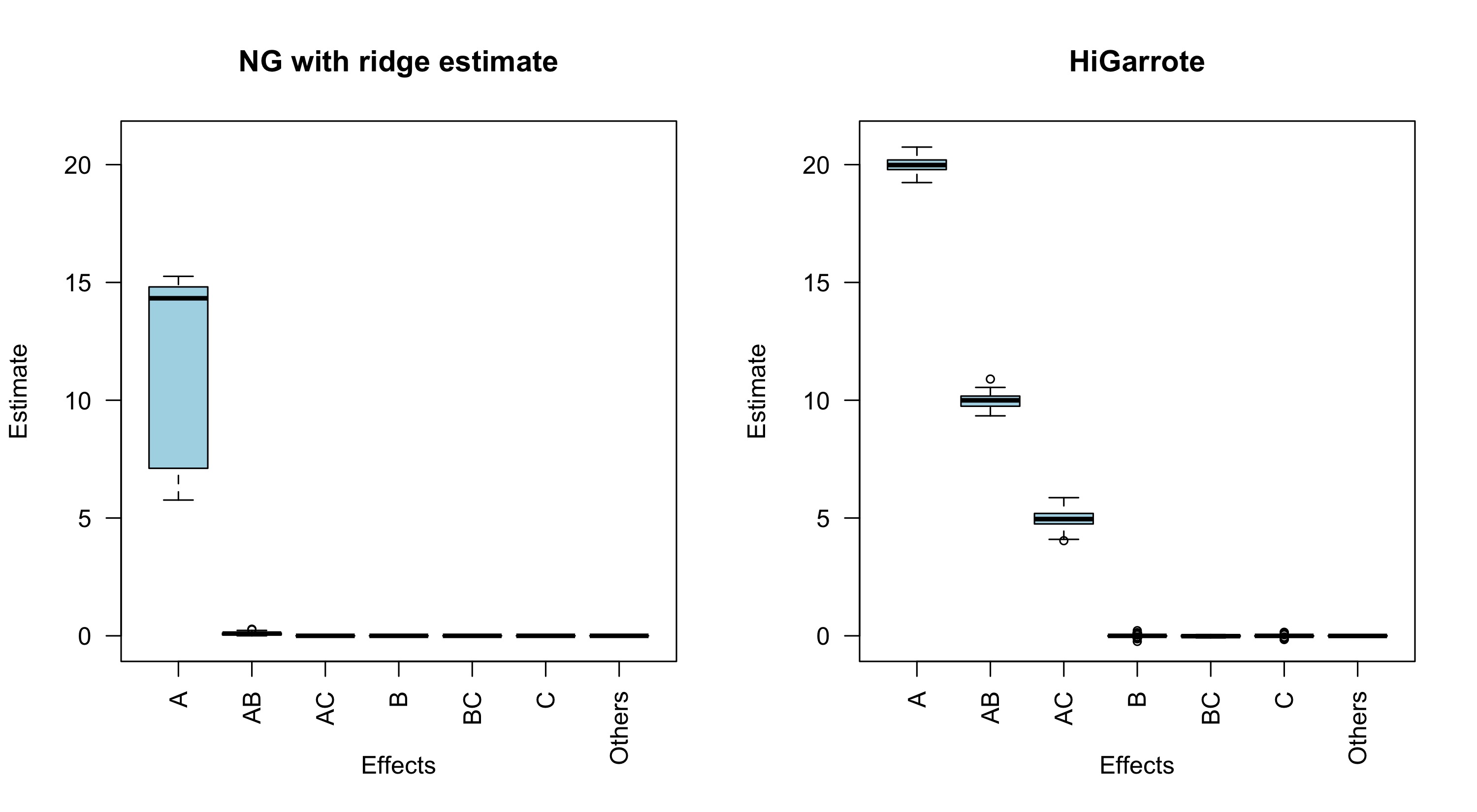} 
    \caption{Comparison of selected effects' estimates using NG with ridge estimate (left) and HiGarrote (right).}
    \label{fig:toy example simulation}
\end{figure}

Figure~\ref{fig:toy example simulation} presents the boxplots of the estimates of selected effects using NG with ridge estimate and HiGarrote.
As shown in Figure~\ref{fig:toy example simulation}, the NG with ridge estimate consistently identifies only the main effect A and occasionally AB.
However, the estimate for A shows large variability, and the estimate for AB remains close to 0.
Importantly, the crucial interaction AC is never selected, indicating its limitation in identifying weaker yet significant interactions.
In contrast, HiGarrote successfully identifies the important effects A, AB, and AC, with estimates close to their true values of 20, 10, and 5.
While HiGarrote occasionally selects additional effects such as B, C, and BC, their estimates are generally very small (near 0) with low variability.

\section{Real Data Examples}
\label{sec:examples}
In this section we evaluate HiGarrote's performance compared to several existing methods using different experimental designs such as regular, nonregular, and screening designs. The following methods are chosen for comparison.\\ 

\noindent {\bf stepH}: This is the stepwise variable selection strategy proposed in \cite{hamada_analysis_1992}.

\noindent {\bf BayesH}: This is the Bayesian variable selection technique proposed in \cite{chipman_bayesian_1997}, which incorporates heredity principles in the SSVS algorithm.

\noindent {\bf hierNet}: This is the LASSO-based method proposed in \cite{bien_lasso_2013}, which is implemented in the R package \texttt{hierNet}.

\noindent {\bf RAMP}: This is the LASSO-based method proposed in \cite{hao_model_2018}, which is implemented in the R package \texttt{RAMP}.

\noindent {\bf GDSARM}: This is the Gauss-Danzig selector aggregation proposed in \citet{singh2024factor}, which is implemented in the R package \texttt{GDSARM}.\\

To make the exposition simple, all the methods are implemented using the weak heredity principle. We are not aware of any software packages that implement stepH and BayesH. Therefore, comparisons with these two methods are limited to the cases available in the literature. The two methods hierNet and RAMP are developed for general-purpose regression problems and therefore, are not suitable for the analysis of certain experimental designs such as mixed-level designs. Therefore, their results are reported only when we are able to implement them in the example. GDSARM can be applied only to two-level designs; so this is implemented only in the example in Section \ref{subsubsec:two-level regular design} and \ref{subsubsec:cast fatigue experiment}. On the other hand, HiGarrote is general and flexible, and can be used with any type of experimental designs.


\subsection{Regular Designs}
\label{subsec:Regular Design}

\subsubsection{Two-Level Regular Design}
\label{subsubsec:two-level regular design}
Consider the $2^{9-5}$ experiment reported by \citet{raghavarao2003heuristic}, where the nine factors are labeled A to J, excluding I.
See Table S1 in the supplementary materials for details of the design matrix and data.
Using half-normal plots and Lenth's method, the aliased effects AH, J, E, G, and CH have smaller p-values while none of them are significant at 0.05 level.
Due to the properties of the design matrix, all the two-factor interactions (2fi) are fully aliased with either main effects or other 2fis.
By ignoring three- and higher-order interactions, we can establish the following aliasing relationships among these five effects: 
$$J = -CF,$$
$$E = -BC,$$
$$G = -AB = -FH,$$
$$AH = BF = DG = EJ,$$
$$CH = GJ =  DE.$$
By applying effect hierarchy principle, we can select J, E, and G from the first three aliasing relationships. Now using effect heredity, the last two aliasing relationships become $DG=EJ$ and $GJ=DE$. These effects cannot be entangled because G, J, and E are active.
Therefore, \citet{raghavarao2003heuristic} conducted a heuristic analysis, and ultimately concluded that the effects J, E, G, EJ, and GJ are active.
Their findings are supported by scientists, who believed that the main effects of E, G, and J, as well as their potential interactions, are likely to be active.


Now consider HiGarrote.
As summarized in Table~\ref{tab:model_summary_2^9-5}, HiGarrote automatically identifies the five active effects without requiring prior knowledge of the aliasing relationships or domain expertise from scientists. This is no accident and the reason behind its selection can be explained as follows. The prior variance of $\beta_{EJ}$ is $.0991\tau^2$, whereas that of $\beta_{DG}$ is $5.3\times 10^{-5}\tau^2$. The observed effect of $EJ=DG$ is divided among the two aliased effects proportional to its prior variance \citep{joseph_bayesian_2006}. Therefore, the initial estimate of $\beta_{EJ}$ is much larger than $\beta_{DG}$, which leads to the automatic selection of EJ by the nonnegative garrote instead of DG. The same thing happens between GJ and DE. Although HiGarrote selects three additional effects: H, HJ, and B, their estimates are relatively low compared to the five active effects.
In terms of computational time tested on a laptop equipped with an Apple M2 chip, HiGarrote takes 0.54 seconds to analyze the experiment.
For ease of comparison, the $R^2$ for all the methods are calculated based on least squares.
We also applied hierNet, RAMP, and GDSARM to this experiment.
However, none of these methods identified any effects.


\begin{table}[ht]
\centering
\caption{Selected Effects in the $2^{9-5}$ Experiment}
\label{tab:model_summary_2^9-5}
\begin{tabular}{llcc}
\toprule
\text{Method} & \text{Model} & \text{$R^2$}\\
\midrule
\citet{raghavarao2003heuristic} & $EJ(-1.70)$, $J(-1.67)$, $E(1.55)$, $G(1.49)$, $GJ(1.39)$ & 70\% \\
HiGarrote & $EJ(-1.29)$, $J(-1.26)$, $E(1.09)$, $G(1.02)$, $GJ(0.87)$, \\
& $H(0.51)$, $HJ(-0.2)$, $B(0.17)$ & 89\% \\
\bottomrule
\end{tabular}
\end{table}

\subsubsection{Mixed-Level Regular Design}
\label{subsubsec:router bit}
\citet{phadke_design_1986} described a 32-run experiment aimed at increasing the lifespan of router bits used in a routing process to cut printed wiring boards from a panel.
The experiment contains seven two-level factors and two four-level qualitative factors which are denoted by A through J with the exclusion of I.
See Table S2 and Table S3 in the supplementary materials for details of factors, design matrix, and data.
For the two qualitative factors, we adopt a convenient coding scheme offered by \citet{wu_experiments_2021}.
The coding scheme generates effects that can be interpreted as the average difference between pairs of levels.
Thus, the coding matrix for factors D and E is
$$\mathbf{U}_j = 
\begin{pmatrix}
    1 & -1 & 1  & -1 \\
    1 & -1 & -1 & 1  \\
    1 & 1  & -1 & -1 \\
    1 & 1  & 1  & 1
\end{pmatrix}.
$$
The main effects of factor D and E are labeled as $D_1$, $D_2$, and $D_3$ and $E_1$, $E_2$, and $E_3$.

This is a regular design and therefore, one can write down all the 15 aliasing relationships. \cite{wu_experiments_2021} used effect hierarchy and heredity principles to identify the following effects: $D_2$, $G$, $J$, $GJ$, and $E_1G=D_2H$. Note that the last aliased effect could not be disentangled because both $D_2$ and $G$ are significant. However, using the physical understanding of the routing process, they argued that $E_1G$ is unlikely to be significant because the four spindles of the routing machine are synchronized. Thus, they finally chose $D_2H$.

On the other hand, as summarized in Table~\ref{tab:model_summary_router_bit}, HiGarrote automatically identifies the five effects without using any physical understanding of the routing process. In addition, it also selects a few more effects $B$, $D_1$, $E_3$, $GH$, $HJ$, $E_3H$. In terms of computational time, HiGarrote takes 6.38 seconds to analyze the experiment.
As mentioned before, since hierNet and RAMP do not generate interactions that are suitable for this experiment, their performance is not reported here.

\begin{table}[ht]
\centering
\caption{Selected Effects in Router Bit Experiment}
\label{tab:model_summary_router_bit}
\begin{tabular}{llcc}
\toprule
\text{Method} & \text{Model} & \text{$R^2$}\\
\midrule
\citet{wu_experiments_2021} & $D_2(2.75)$, $G(-2.56)$, $J(2.31)$, $GJ(-2.25)$, $D_2H(2.25)$ & 62\%\\
HiGarrote & $D_2(2.53)$, $G(-2.33)$, $J(2.06)$, $GJ(-1.98)$, $D_2H(1.97)$, \\
& $GH(1.33)$, $E_3(-1.22)$, $B(-1.14)$,$D_1(.92)$, $HJ(-.86)$,  \\
& $E_3H(-.30)$ & 90\%\\
\bottomrule
\end{tabular}
\end{table}

\subsection{Nonregular Designs}
\label{subsec:Nonregular Design}

\subsubsection{Two-Level Nonregular Design}
\label{subsubsec:cast fatigue experiment}
\citet{hunter_high_1982} used a 12-run Plackett-Burman design to investigate the effects of seven two-level factors on the fatigue life of weld-repaired castings.
The seven factors are denoted by capital letters A through G.
The details of the factors, experimental design, and data are given in  Table S4 and Table S5 of the supplementary materials.
According to the analysis conducted by \citet{hunter_high_1982}, the main effects of factors D and F were found to be significant, but the $R^2$ of the model is only 59\%.
In a very influential work, \citet{hamada_analysis_1992} reanalyzed the experiment using an iterative stepwise regression strategy that incorporates effect heredity (stepH).
They found that by just adding the two-factor interaction FG in the model, $R^2$ increases to 92\%.
They also found that if we remove the main effect D from the model, the $R^2$ only decreases to 89\%, which implies that the effect of D is negligible.
Thus, they claim that effects F and FG are active.



For hierNet where the model is selected by leave-one-out cross-validation, the two important effects F and FG can be successfully identified; however, it also includes too many insignificant effects in the model.
On the other hand, we found that RAMP performs well in this experiment, which selected the same model as the one from stepH.
However, an issue with RAMP is that its algorithm includes quadratic effects such as $A^2$, $B^2$, $\dots$, $G^2$. Since there are only two levels in the experiment, these quadratic effects become the intercept effect which will not be selected, albeit unnecessary computations.
GDSARM, using the settings provided in \citet{singh2024factor}, successfully identified the effects F and FG. However, the algorithm includes several hyperparameters that may require manual tuning for optimal performance. 

HiGarrote identifies five effects: F, FG, D, G, DG, with $R^2 = 96\%$.
Compared to the model selected by stepH, the inclusion of effects D, G, and DG only increases the $R^2$ by 7\%.
Although HiGarrote selects three insignificant effects in the model, the estimates of the effects are low compared to the important effects.
The comparisons are summarized in Table~\ref{tab:model_summary_cast_fatigue}, with the numbers in parentheses indicating the effects estimates. 
The computational time is also reported in the table. HiGarrote is the second slowest among the methods considered in this example, but 0.26 seconds will not be a computational burden for the analyst!

\begin{table}[ht]
\centering
\caption{Selected Effects in Cast Fatigue Experiment}
\label{tab:model_summary_cast_fatigue}
\begin{tabular}{llcc}
\toprule
\text{Method} & \text{Model} & \text{$R^2$} & \text{Runtime}\\
\midrule
\citet{hunter_high_1982} & $F(.46)$, $D(-.26)$ & 59\% & --\\
stepH & $F(.46)$, $FG(-.46)$ & 89\% & .13 sec\\
hierNet & $F(.44)$, $FG(-.26)$, $D(-.17)$, $A(.08)$, \\
& $G(.08)$, $DG(.06)$, $B(.05)$, $BC(-.04)$, \\
& $AE(-.04)$, $AD(-.03)$, $C(-.02)$ & 100\% & .08 sec \\
RAMP & $F(.48)$, $FG(-.5)$ & 89\% & .06 sec\\
GDSARM & $F(.46)$, $FG(-.46)$ & 89\% & .33 sec\\
HiGarrote & $F(.44)$, $FG(-.43)$, $D(-.05)$, $G(.04)$, $DG(.03)$ & 96\% & .26 sec\\
\bottomrule
\end{tabular}
\end{table}

\subsubsection{Mixed-Level Nonregular Design}
\label{subsubsec:blood glucose}
\citet{hamada_analysis_1992} analyzed an 18-run experiment designed to study blood glucose readings of a clinical testing device.
The experiment contains one two-level factor and seven three-level quantitative factors, which are denoted by A through H.
The details of the factors, design matrix, and data are given in Table S6 and Table S7 of the supplementary materials.
Since the experiment contains three-level quantitative factors, we can entertain quadratic terms and their interactions.
The interactions include four components, i.e., linear-by-linear, linear-by-quadratic, quadratic-by-linear, and quadratic-by-quadratic.
So, we consider the search for active effects among 15 main effects (one linear effect, and seven linear and quadratic effects) and 98 interactions.
In the following comparison, the subscripts of $l$ and $q$ indicate the linear and quadratic terms, respectively.

Analysis carried out by \citet{hamada_analysis_1992} using stepH chooses the effects $E_q$, $F_q$, and $E_lF_l$.
The model only achieves an $R^2$ of 68\%.
On the other hand, \citet{chipman_bayesian_1997} using BayesH identified $B_l$, $B_lH_l$, $B_lH_q$, and $B_qH_q$ with $R^2 = 86\%$.
We observed that removing $B_lH_l$ from the model only marginally reduces the $R^2$ to 85\%, suggesting that the effect of $B_lH_l$ is negligible.

It is not ideal to use hierNet and RAMP to analyze the experiment because the way they construct interactions is to cross input variables.
If we treat factors A to H as input variables, the interactions between linear and quadratic effects would not be entertained.
On the other hand, if we treat all linear and quadratic effects as input variables, interactions between effects of the same factor--such as $B_lB_q$--would be included, potentially leading to the selection of cubic effects.
Although treating all linear and quadratic effects as input variables may initially seem reasonable, our analysis revealed that neither hierNet nor RAMP selected any effects.
This suggests that both methods are not suitable for analyzing mixed-level designs.

For HiGarrote, we found that it can identify more significant effects than the existing strategies.
The model selected by HiGarrote contains eight effects, including $B_l$, $B_q$, $F_l$, $H_l$, $H_q$, $B_lH_q$, $B_qH_l$, $B_qH_q$, with an $R^2$ of 97\%.
In the selected model, all the effects except $H_l$ and $H_q$ are significant.
In terms of computational time, it takes 1.40 seconds for HiGarrote to analyze the experiment.
The results are summarized in Table~\ref{tab:model_summary_blood_glucose}.

\begin{table}[ht]
\centering
\caption{Selected Effects in Blood Glucose Experiment}
\label{tab:model_summary_blood_glucose}
\begin{tabular}{llcc}
\toprule
\text{Method} & \text{Model} & \text{$R^2$}\\
\midrule
stepH & $E_lF_l(-5.52)$, $E_q(4.33)$, $F_q(4.02)$ & 68\%\\
BayesH & $B_lH_q(6.64)$, $B_qH_q(-5.43)$, $B_l(-2.85)$, $B_lH_l(.71)$ & 86\%\\
HiGarrote & $B_lH_q(6.52)$, $B_qH_q(-5.10)$, $B_l(-2.60)$, $B_q(1.28)$, \\
& $B_qH_l(.99)$, $H_l(-.45)$, $F_l(-.34)$, $H_q(-.05)$ & 97\% \\
\bottomrule
\end{tabular}
\end{table}

\subsection{Screening Designs}
\label{subsec:screening design}

\subsubsection{Definitive Screening Design}
\label{subsubsec:DSD}
\citet{jones2021novel} described a 21-run experiment aimed at creating a special resin for car vents designed to prevent moisture accumulation within a car's housing.
The experiment is carried out using a definitive screening design (DSD) with nine continuous factors, labeled A through J, with the exclusion of I.
Details of factors and dataset are provided in Table S8 and Table S9 of the supplementary materials. The response is taken as the logarithm of Impurity.



According to the analysis carried out by \citet{jones2021novel}, the linear effects A and F, and their interaction AF are identified as significant with an $R^2$ of 93\%.
In addition to the three effects selected by \citet{jones2021novel}, HiGarrote identifies five more effects, among which four of them are significant.
The $R^2$ increases to 99\%.
In terms of computational time, it takes 1.24 seconds for HiGarrote to analyze the experiment.
Since both hierNet and RAMP can handle the full quadratic model, their performance can be examined here.
It turns out that hierNet does not perform well in this experiment because there are too many effects being selected.
In total, the selected model contains 16 effects comprised of 7 linear effects and 9 interactions.
On the other hand, RAMP only identifies F.
Since F is highly important, RAMP's model is good with an $R^2$ of 88\%.
The results are summarized in Table~\ref{tab:model_summary_resin}.
To make the table concise, the selected model of hierNet is not reported.

\begin{table}[ht]
\centering
\caption{Selected Effects in Resin Experiment}
\label{tab:model_summary_resin}
\begin{tabular}{llcc}
\toprule
\text{Method} & \text{Model} & \text{$R^2$}\\
\midrule
\citet{jones2021novel} & $F(-2.21)$, $A(.47)$, $AF(.23)$ & 93\%\\
RAMP & $F(-2.21)$ & 88\%\\
HiGarrote & $F(-2.20)$, $A(.43)$, $FJ(-.30)$, $E(-.23)$, \\
& $BF(.16)$, $AJ(-.10)$, $AF(.05)$, $F^2(.04)$ & 99\%\\
\bottomrule
\end{tabular}
\end{table}

\subsubsection{Supersaturated Design}
\label{subsubsec:supersaturated design}
\citet{lin_new_1993} employed a half-fraction of a 28-run PB design to investigate an experiment aimed at developing an epoxy adhesive system, as described by \citet{williams_designed_1968}.
The original design matrix contains 14 runs and 24 factors.
Since factors 13 and 16 were assigned to the same columns in the original design matrix, only factor 13 is listed here.
The design matrix and data are given in Table S10 of the supplementary materials.
Since supersaturated designs are generally used for factor screening, we focus only on the main effects.
Therefore, we will exclude hierNet, RAMP, and GDSARM in this example, as these methods are designed to identify models that encompass quadratic effects and two-factor interactions. Although HiGarrote is also intended for models with hierarchical effects, we apply it in this example to show its generality. The results are summarized in Table~\ref{tab:model_summary_epoxy}. We can see that HiGarrote identifies the same set of factors as in \citet{lin_new_1993} and \cite{chipman_bayesian_1997}. In terms of computational time, HiGarrote takes about 5.80 seconds to perform the analysis, in which most of the time is spent on optimizing the 24 hyperparameters of the correlation matrix and the noise ratio.


\begin{table}[ht]
\centering
\caption{Selected Effects in Epoxy Experiment}
\label{tab:model_summary_epoxy}
\begin{tabular}{llcc}
\toprule
\text{Method} & \text{Model} & \text{$R^2$}\\
\midrule
\citet{lin_new_1993} & $15 (-71.26)$, $20 (-27.98)$, $12 (-26.77)$, $4 (20.73)$, $10 (-9.40)$ & 97\%\\
BayesH & $15 (-70.48)$, $20 (-29.20)$, $12 (-25.29)$, $4 (22.12)$ & 95\%\\
HiGarrote & $15 (-61.22)$, $12 (-25.84)$, $20 (-22.19)$, $10 (-8.42)$, $4 (1.29)$ & 97\%\\
\bottomrule
\end{tabular}
\end{table}

\section{Conclusion}
\label{sec:conc}

Regression analysis of experimental data requires special care because of two main reasons: (i) the data size is small and (ii) there is a need to estimate higher order effects that satisfy effect hierarchy and heredity principles. We have proposed a modified version of nonnegative garrote method, called HiGarrote, that incorporates the hierarchical relationships among the effects. A major innovation in this new technique is to carefully devise an approach to obtain an initial estimate for the nonnegative garrote. For this purpose we used generalized ridge regression with the ridge parameters chosen based on a Gaussian process prior on the underlying function. The main computational cost of our method comes from the fitting of the Gaussian process, which increases at the rate of $\mathcal{O}(n^3)$, where $n$ is the size of the data. However, since $n$ is small in experiments, this is not of much concern. 

We have applied our method to many different kinds of experiments report in the literature and found that it works well. The most attractive feature of the method is that it is very general and requires no manual tuning. Thus, it can be used to automate the analysis of experiments. We plan to provide the implementation of the code in an R package. Some practitioners may still prefer to analyze their experiments manually. However, even in such cases, HiGarrote could be tried so that the accuracy of their analyses can be quickly verified.

Another novelty of the proposed method is that it makes a connection to the Gaussian process modeling which is widely applied to the analysis of computer experiments. Practitioners have so far stayed away from Gaussian process regression in physical experiments because they felt that such a complex nonparameteric regression model is unnecessary to analyze the small data from  experiments, which is also corrupted by noise. Nevertheless, we found Gaussian process modeling to be useful, at least to get a good initial estimate for the linear regression parameters. One may think then why not just use Gaussian process regression in the analysis of physical experiments and why we still need to stick with linear regression. The reason is that linear regression does a better job in identifying significant effects that are interpretable, whereas variable selection with Gaussian process regression is not trivial \citep{linkletter2006variable}. However, some recent advances in global sensitivity analysis \citep{huang2024factor} offer some promising alternatives in this direction, which needs more investigation.

\vspace{.25in}
\begin{appendices}
{\Large \bfseries \noindent Appendix: Proof of ~\eqref{eq:prior for qualitative factor}}


Suppose that there are $p$ qualitative three-level factors and their model matrices are encoded according to Helmert coding.
Then, the model matrix for factor $j$ is
$$\mathbf{U}_j = 
\begin{pmatrix}
    1 & -\sqrt{\frac{3}{2}} & - \sqrt{\frac{1}{2}} \\
    1 & \sqrt{\frac{3}{2}} & -\sqrt{\frac{1}{2}} \\
    1 & 0 & 2\sqrt{\frac{1}{2}} \\
\end{pmatrix}.
$$
Given that each factor is represented by two dummy variables, and based on the model matrix, these dummy variables are expressed as $m_{j_1} = (-1,1,0)'$ and $m_{j_2} = (-1,-1,2)'$.
Then, the two correlation matrices are
$$
\mathbf{\Psi}_{j_1} = 
\begin{pmatrix}
    1 & \rho_{j_1} & \rho_{j_1} \\
    \rho_{j_1} & 1 & \rho_{j_1} \\
    \rho_{j_1} & \rho_{j_1} & 1 
\end{pmatrix}, \;
\mathbf{\Psi}_{j_2} = 
\begin{pmatrix}
    1 & 1 & \rho_{j_2} \\
    1 & 1 & \rho_{j_2} \\
    \rho_{j_2} & \rho_{j_2} & 1 
\end{pmatrix}.
$$
According to the product correlation structure, the correlation matrix of factor $j$ is
$$
\mathbf{\Psi}_j = \mathbf{\Psi}_{j_1} \odot \mathbf{\Psi}_{j_2} = 
\begin{pmatrix}
    1 & \rho_{j_1} & \rho_{j_1} \rho_{j_2} \\
    \rho_{j_1} & 1 & \rho_{j_1} \rho_{j_2} \\
    \rho_{j_1} \rho_{j_2} & \rho_{j_1} \rho_{j_2} & 1
\end{pmatrix},
$$
where $\odot$ is the Hadamard product.
Then,
$$\mathbf{U}_j^{-1} \mathbf{\Psi}_j (\mathbf{U}_j^{-1})' = \frac{1}{9}
\begin{pmatrix}
    3+2\rho_{j1}+4\rho_{j1}\rho_{j2} & 0 & \sqrt{2}(-\rho_{j1} + \rho_{j1}\rho_{j2}) \\
    0 & 3(1-\rho_{j1}) & 0 \\
    \sqrt{2}(-\rho_{j1} + \rho_{j1}\rho_{j2}) & 0 & 3+\rho_{j1}-4\rho_{j1}\rho_{j2}
\end{pmatrix}.
$$
Now using (\ref{eq:kronecker}), the marginal distributions of $\beta_i$'s can be obtained from (\ref{eq:general prior}). This gives ~\eqref{eq:prior for qualitative factor}.


\end{appendices}

\vspace{.25in}
\noindent{\Large\bf Supplementary Materials}

\noindent The file contains the details of the factors and their levels, design matrix, and data for the examples in Section 4.

\vspace{.25in}
\noindent{\Large\bf Disclosure of Interest}

\noindent No potential conflict of interest was reported by the authors.

\vspace{.25in}
\noindent{\Large\bf Data Availability Statement}

\noindent The data are provided in the supplementary materials.

\vspace{.25in}
\noindent{\Large\bf Funding }

\noindent This research is supported by U.S. National Science Foundation grants DMS-2310637.
\vspace{.25in}

\bibliographystyle{apalike}
\bibliography{bibiliography}

\end{document}


\begin{center}
    {\Large\bf Supplementary Materials}\\ 
    {\large\bf Automated Analysis of Experiments using Hierarchical Garrote}
\end{center}

\section{Two-Level Regular Design}
\begin{table}[h]
\centering
\caption{The $2^{9-5}$ Design Matrix and Data}
\label{tab:2^9-5 design Matrix and Data}
\begin{tabular}{ccccccccccc}
\toprule
Run & A & B & C & D & E & F & G & H & J & Response \\
\midrule
1 & - & - & - & - & - & - & - & - & - & 136.475 \\
2 & + & + & - & + & + & - & - & - & - & 147.775 \\
3 & + & - & - & + & - & - & + & + & - & 142.425 \\
4 & + & - & + & + & + & - & + & + & + & 141.800 \\
5 & + & + & + & + & - & - & - & - & + & 136.675 \\
6 & - & + & - & - & + & - & + & + & - & 150.725 \\
7 & - & + & + & - & - & - & + & + & + & 142.800 \\
8 & - & - & + & - & + & - & - & - & + & 135.825 \\
9 & + & + & + & - & - & + & - & + & - & 143.476 \\
10 & - & + & - & + & + & + & + & - & + & 145.150 \\
11 & + & - & + & - & + & + & + & - & - & 142.600 \\
12 & - & - & - & + & - & + & - & + & + & 139.375 \\
13 & + & + & - & - & + & + & - & + & + & 139.650 \\
14 & + & - & - & - & - & + & + & - & + & 144.775 \\
15 & - & - & + & + & + & + & - & + & - & 148.275 \\
16 & - & + & + & + & - & + & + & - & - & 141.075 \\
\bottomrule
\end{tabular}
\end{table}

\section{Mixed-Level Regular Design}
\begin{table}[h]
\centering
\caption{Factors and Levels for Router Bit Experiment}
\label{tab:Factors and Levels for Router Bit Experiment}
\begin{tabular}{clp{1.5cm}p{1.5cm}p{1.5cm}p{1.5cm}}
\toprule
\multicolumn{2}{c}{\text{Factor}} & \multicolumn{4}{c}{\text{Level}} \\
\cmidrule(lr){3-6}
& & \text{1} & \text{2} & \text{3} & \text{4} \\
\midrule
\text{A}, & Suction (in of Hg)      & 1       & 2       &         &         \\
\text{B}, & x--y feed (in/min)      & 60      & 80      &         &         \\
\text{C}, & In-feed (in/min)        & 10      & 50      &         &         \\
\text{D}, & Bit type                & 1       & 2       & 3       & 4       \\
\text{E}, & Spindle position        & 1       & 2       & 3       & 4       \\
\text{F}, & Suction foot            & SR      & BB      &         &         \\
\text{G}, & Stacking height (in)    & 3/16    & 1/4     &         &         \\
\text{H}, & Slot depth (mil)        & 60      & 100     &         &         \\
\text{J}, & Speed (rpm)             & 30,000  & 40,000  &         &         \\
\bottomrule
\end{tabular}
\end{table}

\begin{table}
\centering
\caption{Design Matrix and Data for Router Bit Experiment}
\label{tab:router_bit_experiment}
\begin{tabular}{ccccccccccc}
\toprule
Run & A & B & C & D & E & F & G & H & J & Lifetime \\ 
\midrule
1  & $-$ & $-$ & $-$ & 1 & 1 & $-$ & $-$ & $-$ & $-$ & 3.5 \\
2  & $-$ & $-$ & $-$ & 2 & 2 & $+$ & $+$ & $-$ & $-$ & 0.5 \\
3  & $-$ & $-$ & $-$ & 3 & 4 & $-$ & $+$ & $+$ & $-$ & 0.5 \\
4  & $-$ & $-$ & $-$ & 4 & 3 & $+$ & $-$ & $+$ & $-$ & 17.0 \\
5  & $-$ & $+$ & $+$ & 3 & 1 & $+$ & $+$ & $-$ & $-$ & 0.5 \\
6  & $-$ & $+$ & $+$ & 4 & 2 & $-$ & $-$ & $-$ & $-$ & 2.5 \\
7  & $-$ & $+$ & $+$ & 1 & 4 & $+$ & $-$ & $+$ & $-$ & 0.5 \\
8  & $-$ & $+$ & $+$ & 2 & 3 & $-$ & $+$ & $+$ & $-$ & 0.5 \\
9  & $+$ & $-$ & $+$ & 4 & 1 & $-$ & $+$ & $+$ & $-$ & 17.0 \\
10 & $+$ & $-$ & $+$ & 3 & 2 & $+$ & $-$ & $+$ & $-$ & 2.5 \\
11 & $+$ & $-$ & $+$ & 2 & 4 & $-$ & $-$ & $-$ & $-$ & 0.5 \\
12 & $+$ & $-$ & $+$ & 1 & 3 & $+$ & $+$ & $-$ & $-$ & 3.5 \\
13 & $+$ & $+$ & $-$ & 2 & 1 & $+$ & $-$ & $+$ & $-$ & 0.5 \\
14 & $+$ & $+$ & $-$ & 1 & 2 & $-$ & $+$ & $+$ & $-$ & 2.5 \\
15 & $+$ & $+$ & $-$ & 4 & 4 & $+$ & $+$ & $-$ & $-$ & 0.5 \\
16 & $+$ & $+$ & $-$ & 3 & 3 & $-$ & $-$ & $-$ & $-$ & 3.5 \\
17 & $-$ & $-$ & $-$ & 1 & 1 & $-$ & $-$ & $-$ & $+$ & 17.0 \\
18 & $-$ & $-$ & $-$ & 2 & 2 & $+$ & $+$ & $-$ & $+$ & 0.5 \\
19 & $-$ & $-$ & $-$ & 3 & 4 & $-$ & $+$ & $+$ & $+$ & 0.5 \\
20 & $-$ & $-$ & $-$ & 4 & 3 & $+$ & $-$ & $+$ & $+$ & 17.0 \\
21 & $-$ & $+$ & $+$ & 3 & 1 & $+$ & $+$ & $-$ & $+$ & 0.5 \\
22 & $-$ & $+$ & $+$ & 4 & 2 & $-$ & $-$ & $-$ & $+$ & 17.0 \\
23 & $-$ & $+$ & $+$ & 1 & 4 & $+$ & $-$ & $+$ & $+$ & 14.5 \\
24 & $-$ & $+$ & $+$ & 2 & 3 & $-$ & $+$ & $+$ & $+$ & 0.5 \\
25 & $+$ & $-$ & $+$ & 4 & 1 & $-$ & $+$ & $+$ & $+$ & 17.0 \\
26 & $+$ & $-$ & $+$ & 3 & 2 & $+$ & $-$ & $+$ & $+$ & 3.5 \\
27 & $+$ & $-$ & $+$ & 2 & 4 & $-$ & $-$ & $-$ & $+$ & 17.0 \\
28 & $+$ & $-$ & $+$ & 1 & 3 & $+$ & $+$ & $-$ & $+$ & 3.5 \\
29 & $+$ & $+$ & $-$ & 2 & 1 & $+$ & $-$ & $+$ & $+$ & 0.5 \\
30 & $+$ & $+$ & $-$ & 1 & 2 & $-$ & $+$ & $+$ & $+$ & 3.5 \\
31 & $+$ & $+$ & $-$ & 4 & 4 & $+$ & $+$ & $-$ & $+$ & 0.5 \\
32 & $+$ & $+$ & $-$ & 3 & 3 & $-$ & $-$ & $-$ & $+$ & 17.0 \\
\bottomrule
\end{tabular}
\end{table}

\newpage
\section{Two-Level Nonregular Design}
\begin{table}[h]
    \centering
    \caption{Factors and Levels for Cast Fatigue Experiment}
    \label{tab:Factors and Levels for Cast Fatigue Experiment}
    \begin{tabular}{clcc}
        \toprule
        \multicolumn{2}{c}{\text{Factor}} & \multicolumn{2}{c}{\text{Level}} \\
        \cmidrule(lr){3-4}
        & & 1 & $-1$ \\
        \midrule
        A, & initial structure & $\beta$ treat & as received \\
        B, & bead size & large & small \\
        C, & pressure treat & HIP & none \\
        D, & heat treat & solution treat/age & anneal \\
        E, & cooling rate & rapid & slow \\
        F, & polish & mechanical & chemical \\
        G, & final treat & peen & none \\
        \bottomrule
    \end{tabular}
\end{table}

\begin{table}[h]
\centering
\caption{Design Matrix and Data for Cast Fatigue Experiment}
\label{tab:cast fatigue}
\begin{tabular}{crrrrrrrc}
\toprule
Run & A  & B  & C  & D  & E  & F  & G  & Response    \\ 
\midrule
1  & 1  & 1  & -1 & 1  & 1  & 1  & -1 & 6.058 \\
2  & 1  & -1 & 1  & 1  & 1  & -1 & -1 & 4.733 \\
3  & -1 & 1  & 1  & 1  & -1 & -1 & -1 & 4.625 \\
4  & 1  & 1  & 1  & -1 & -1 & -1 & 1  & 5.899 \\
5  & 1  & 1  & -1 & -1 & -1 & 1  & -1 & 7.000 \\
6  & 1  & -1 & -1 & -1 & 1  & -1 & 1  & 5.752 \\
7  & -1 & -1 & -1 & 1  & -1 & 1  & 1  & 5.682 \\
8  & -1 & -1 & 1  & -1 & 1  & 1  & -1 & 6.607 \\
9  & -1 & 1  & -1 & 1  & 1  & -1 & 1  & 5.818 \\
10 & 1  & -1 & 1  & 1  & -1 & 1  & 1  & 5.917 \\
11 & -1 & 1  & 1  & -1 & 1  & 1  & 1  & 5.863 \\
12 & -1 & -1 & -1 & -1 & -1 & -1 & -1 & 4.809 \\
\bottomrule
\end{tabular}
\end{table}

\newpage
\section{Mixed-Level Nonregular Design}
\begin{table}[h]
\centering
\caption{Factors and Levels for Blood Glucose Experiment}
\label{tab:Factors and Levels for Blood Glucose Experiment}
\begin{tabular}{cllcc}  
\toprule
\multicolumn{2}{c}{\text{Factor}} & \multicolumn{3}{c}{\text{Level}} \\ 
\cmidrule(lr){3-5}
 & & \text{1} & \text{2} & \text{3} \\
\midrule
\text{A},& Wash                     & No      & Yes     &  \\
\text{B},& Microvial volume (ml)     & 2.0     & 2.5     & 3.0  \\
\text{C},& Caras H$_2$O level (ml)   & 20      & 28      & 35  \\
\text{D},& Centrifuge rpm            & 2,100   & 2,300   & 2,500  \\
\text{E},& Centrifuge time (min)     & 1.75    & 3       & 4.5  \\
\text{F},& Sensitivity, absorption   & (.10, 2.5)  & (.25, 2)  & (.50, 1.5)  \\
\text{G},& Temperature ($^\circ$C)   & 25      & 30      & 37  \\
\text{H},& Dilution ratio            & 1:51    & 1:101   & 1:151  \\
\bottomrule
\end{tabular}
\end{table}

\begin{table}[h]
\centering
\caption{Design Matrix and Data, Blood Glucose Experiment}
\renewcommand{\arraystretch}{0.9} 
\label{tab:blood_glucose}
\begin{tabular}{ccccccccccc}
    \toprule
    \multirow{2}{*}{\text{Run}} & \multicolumn{8}{c}{\text{Factor}} & \multirow{2}{*}{\text{Mean}} \\
    \cmidrule(r){2-9}
    \text{} & \text{A} & \text{G} & \text{B} & \text{C} & \text{D} & \text{E} & \text{F} & \text{H} & \text{reading} \\
    \midrule
        1  & 1 & 1 & 1 & 1 & 1 & 1 & 1 & 1 & 97.94 \\
        2  & 1 & 1 & 2 & 2 & 2 & 2 & 2 & 2 & 83.40 \\
        3  & 1 & 1 & 3 & 3 & 3 & 3 & 3 & 3 & 95.88 \\
        4  & 1 & 2 & 1 & 1 & 2 & 2 & 3 & 3 & 88.86 \\
        5  & 1 & 2 & 2 & 2 & 3 & 3 & 1 & 1 & 106.58 \\
        6  & 1 & 2 & 3 & 3 & 1 & 1 & 2 & 2 & 89.57 \\
        7  & 1 & 3 & 1 & 2 & 1 & 3 & 2 & 3 & 91.98 \\
        8  & 1 & 3 & 2 & 3 & 2 & 1 & 3 & 1 & 98.41 \\
        9  & 1 & 3 & 3 & 1 & 3 & 2 & 1 & 2 & 87.56 \\
        10 & 2 & 1 & 1 & 3 & 3 & 2 & 2 & 1 & 88.11 \\
        11 & 2 & 1 & 2 & 1 & 1 & 3 & 3 & 2 & 83.81 \\
        12 & 2 & 1 & 3 & 2 & 2 & 1 & 1 & 3 & 98.27 \\
        13 & 2 & 2 & 1 & 2 & 3 & 1 & 3 & 2 & 115.52 \\
        14 & 2 & 2 & 2 & 3 & 1 & 2 & 1 & 3 & 94.89 \\
        15 & 2 & 2 & 3 & 1 & 2 & 3 & 2 & 1 & 94.70 \\
        16 & 2 & 3 & 1 & 3 & 2 & 3 & 1 & 2 & 121.62 \\
        17 & 2 & 3 & 2 & 1 & 3 & 1 & 2 & 3 & 93.86 \\
        18 & 2 & 3 & 3 & 2 & 1 & 2 & 3 & 1 & 96.10 \\
    \bottomrule
\end{tabular}
\end{table}

\newpage
\section{Definitive Screening Design}
\begin{table}[h]
\centering
\caption{Factors and Responses for the Process Steps}
\label{tab:Factors and Responses for the Process Steps}
\renewcommand{\arraystretch}{0.5} 
\begin{tabular}{p{0.5cm} p{3cm} p{3cm}}
\toprule
& \multicolumn{1}{c}{Polymerization Step} &  \multicolumn{1}{c}{Finish Step}\\
\midrule
& \multicolumn{2}{c}{\textbf{Factors}} \\
A, & \multicolumn{1}{c}{Ingredient 1} & \multicolumn{1}{c}{Ingredient 1} \\
B, & \multicolumn{1}{c}{Ingredient 2} & \multicolumn{1}{c}{Ingredient 2} \\
C, & \multicolumn{1}{c}{Process Condition} & \multicolumn{1}{c}{Process Condition} \\
D, &  & \multicolumn{1}{c}{Amount NA} \\
E, &  & \multicolumn{1}{c}{Amount AW} \\
F, &  & \multicolumn{1}{c}{N Wash Cycles} \\
G, &  & \multicolumn{1}{c}{Wash Temp}  \\
H, &  & \multicolumn{1}{c}{Dry Soak}  \\
J, &  & \multicolumn{1}{c}{Dry Time}  \\
\midrule
& \multicolumn{2}{c}{\textbf{Responses}} \\
& \multicolumn{1}{c}{MFI} & \multicolumn{1}{c}{Impurity} \\
& \multicolumn{1}{c}{TGA} & \\
\bottomrule
\end{tabular}
\end{table}

\begin{table}[h]
\centering
\caption{Design Matrix and Data for Resin Experiment}
\label{tab:resin_experiment}
\renewcommand{\arraystretch}{0.7} 
\begin{tabular}{ccccccccccccc}
\toprule
    Run & A & B & C & D & E & F & G & H & J & MFI & TGA & Impurity \\

\midrule
    1 & 6.5 & 1.64 & 62 & 3.2 & 3.6 & 8 & 50 & 0 & 24 & 0.27 & 429 & 1 \\
    2 & 6.5 & 0 & 62 & 5.2 & 5.1 & 3 & 50 & 0 & 18 & 0.21 & 430.5 & 200 \\
    3 & 9 & 0 & 70 & 5.2 & 3.6 & 3 & 30 & 6 & 12 & 4.3 & 413.3 & 260 \\
    4 & 6.5 & 0 & 70 & 3.2 & 5.1 & 3 & 40 & 6 & 24 & 4.17 & 413.4 & 190 \\
    5 & 6.5 & 1.64 & 66 & 5.2 & 2.1 & 3 & 50 & 6 & 12 & 0.87 & 425.9 & 88 \\
    6 & 7.75 & 1.64 & 70 & 5.2 & 5.1 & 8 & 50 & 6 & 24 & 4.01 & 420.4 & 1 \\
    7 & 6.5 & 0 & 62 & 5.2 & 2.1 & 8 & 30 & 6 & 24 & 0.23 & 428.1 & 1 \\
    8 & 7.75 & 0 & 62 & 3.2 & 2.1 & 3 & 30 & 0 & 12 & 0.26 & 419.7 & 160 \\
    9 & 9 & 1.64 & 62 & 5.2 & 5.1 & 3 & 30 & 3 & 24 & 0.25 & 427 & 200 \\
    10 & 6.5 & 1.64 & 62 & 3.2 & 5.1 & 5.5 & 30 & 6 & 12 & 0.34 & 428.7 & 3.6 \\
    11 & 7.75 & 0.82 & 66 & 4.2 & 3.6 & 5.5 & 40 & 3 & 18 & 0.81 & 426.4 & 12 \\
    12 & 9 & 1.64 & 70 & 3.2 & 5.1 & 3 & 50 & 0 & 12 & 3.45 & 423 & 160 \\
    13 & 6.5 & 0.82 & 70 & 5.2 & 5.1 & 8 & 30 & 0 & 12 & 3.47 & 417.1 & 1 \\
    14 & 9 & 0.82 & 62 & 3.2 & 2.1 & 3 & 50 & 6 & 24 & 0.24 & 426.1 & 390 \\
    15 & 9 & 1.64 & 62 & 5.2 & 2.1 & 8 & 40 & 0 & 12 & 0.22 & 426.2 & 23 \\
    16 & 6.5 & 0 & 70 & 3.2 & 2.1 & 8 & 50 & 3 & 12 & 3.67 & 416.2 & 1 \\
    17 & 9 & 0 & 62 & 4.2 & 5.1 & 8 & 50 & 6 & 12 & 0.21 & 426.7 & 2.6 \\
    18 & 9 & 0 & 70 & 5.2 & 2.1 & 5.5 & 50 & 0 & 24 & 5.33 & 411.9 & 22 \\
    19 & 6.5 & 1.64 & 70 & 4.2 & 2.1 & 3 & 30 & 0 & 24 & 4.22 & 418.2 & 230 \\
    20 & 9 & 1.64 & 70 & 3.2 & 2.1 & 8 & 30 & 6 & 18 & 3.94 & 421.9 & 4.5 \\
    21 & 9 & 0 & 66 & 3.2 & 5.1 & 8 & 30 & 0 & 24 & 1 & 421.9 & 1 \\
\bottomrule
\end{tabular}
\end{table}

\section{Supersaturated Design}
\newpage
\begin{sidewaystable}
\centering
\caption{Design Matrix and Data for Epoxy Experiment}
\label{tab:Design Matrix and Data for Epoxy Experiment}
\setlength\tabcolsep{3pt} 
\begin{tabular}{crrrrrrrrrrrrrrrrrrrrrrrc}
\toprule
Run & 1 & 2 & 3 & 4 & 5 & 6 & 7 & 8 & 9 & 10 & 11 & 12 & 13 & 14 & 15 & 17 & 18 & 19 & 20 & 21 & 22 & 23 & 24 & Response \\ 
\midrule
1  & 1  & 1  & 1  & -1 & -1 & -1 & 1  & 1  & 1  & 1  & 1  & -1 & 1  & -1 & -1 & 1  & -1 & -1 & 1  & -1 & -1 & -1 & 1  & 133 \\
2  & 1  & -1 & -1 & -1 & -1 & -1 & 1  & 1  & 1  & -1 & -1 & -1 & 1  & 1  & 1  & -1 & 1  & -1 & -1 & 1  & 1  & -1 & -1 & 62  \\
3  & 1  & 1  & -1 & 1  & 1  & -1 & -1 & -1 & -1 & 1  & -1 & 1  & 1  & 1  & 1  & 1  & -1 & -1 & -1 & -1 & 1  & 1  & -1 & 45  \\
4  & 1  & 1  & -1 & 1  & -1 & 1  & -1 & -1 & -1 & 1  & 1  & -1 & 1  & -1 & 1  & -1 & 1  & 1  & 1  & -1 & -1 & -1 & -1 & 52  \\
5  & -1 & -1 & 1  & 1  & 1  & 1  & -1 & 1  & 1  & -1 & -1 & -1 & 1  & -1 & 1  & 1  & -1 & -1 & 1  & -1 & 1  & 1  & 1  & 56  \\
6  & -1 & -1 & 1  & 1  & 1  & 1  & 1  & -1 & 1  & 1  & 1  & -1 & -1 & 1  & 1  & 1  & 1  & 1  & 1  & 1  & 1  & -1 & -1 & 47  \\
7  & -1 & -1 & -1 & -1 & 1  & -1 & -1 & 1  & -1 & 1  & -1 & 1  & 1  & 1  & -1 & 1  & 1  & 1  & 1  & 1  & -1 & -1 & 1  & 88  \\
8  & -1 & 1  & 1  & -1 & -1 & 1  & -1 & 1  & -1 & 1  & -1 & -1 & -1 & -1 & -1 & -1 & -1 & 1  & -1 & 1  & 1  & 1  & -1 & 193 \\
9  & -1 & -1 & -1 & -1 & -1 & 1  & 1  & -1 & -1 & -1 & 1  & 1  & -1 & -1 & 1  & 1  & 1  & -1 & -1 & -1 & -1 & 1  & 1  & 32  \\
10 & 1  & 1  & 1  & 1  & -1 & 1  & 1  & 1  & -1 & -1 & -1 & 1  & -1 & 1  & 1  & 1  & -1 & 1  & -1 & 1  & -1 & -1 & 1  & 53  \\
11 & -1 & 1  & -1 & 1  & 1  & -1 & -1 & 1  & 1  & -1 & 1  & -1 & -1 & 1  & -1 & -1 & 1  & 1  & -1 & -1 & -1 & 1  & 1  & 276 \\
12 & 1  & -1 & -1 & -1 & 1  & 1  & 1  & -1 & 1  & 1  & 1  & 1  & 1  & -1 & -1 & -1 & -1 & 1  & -1 & 1  & 1  & 1  & 1  & 145 \\
13 & 1  & 1  & 1  & 1  & 1  & -1 & 1  & -1 & 1  & -1 & -1 & 1  & -1 & -1 & -1 & -1 & 1  & -1 & 1  & 1  & -1 & 1  & -1 & 130 \\
14 & -1 & -1 & 1  & -1 & -1 & -1 & -1 & -1 & -1 & -1 & 1  & 1  & -1 & 1  & -1 & -1 & -1 & -1 & 1  & -1 & 1  & -1  & -1 & 127 \\
\bottomrule
\end{tabular}
\end{sidewaystable}